\documentclass[letterpaper]{jpconf}
\usepackage{graphicx}
\usepackage{amssymb}
\begin{document}

\title{MSW Oscillations - LMA and Subdominant Effects}

\author{Alexander Friedland}

\address{Theoretical Division, T-8, MS B285, Los Alamos National
Laboratory, Los Alamos, NM 87545}

\ead{friedland@lanl.gov}

\begin{abstract}
  New physics near the TeV scale could modify neutrino-matter
  interactions or generate a relatively large neutrino magnetic
  (transition) moment.  Both types of effects have been discussed
  since the 1970's as alternatives to mass-induced neutrino flavor
  oscillations. Nowadays, the availability of high-statistics
  data makes it possible to turn the idea around and ask: How well
  do the simple mass-induced oscillations describe solar neutrinos? At
  what level are the above-mentioned nonstandard effects excluded? Can
  we use solar neutrinos to constrain physics beyond the Standard
  Model?
  These notes review the sensitivity of the present-day solar neutrino
  experiments to the nonstandard neutrino interactions and transition
  moment and outline progress that may be expected in the near future.
  Based on a talk given at the Neutrino 2006 conference
  \cite{Neutrino06}.
\end{abstract}

Preprint: LA-UR-06-6774

\section{Standard LMA solution: basic features}
\label{sect:stdLMA}


The most basic experimental fact about the neutrinos from the Sun is
that the electron neutrino survival probability, $P_{ee}^{std}\equiv
P(\nu_e\rightarrow\nu_e)$, is measured to vary as a function of the
neutrino energy. At the high end of the spectrum ($E_\nu\gtrsim 6-7$
MeV) the SNO \cite{SNO2005} and Super-Kamiokande \cite{SK2003}
experiments have established that $P_{ee}^{std}$ is about $\sim
34\pm3$\%. The gallium experiments \cite{GNO2000}, however, which are
sensitive to both high- and low-energy neutrinos, see a higher
survival probability: the measured rate is $74\pm7$ SNU, whereas the
standard solar model prediction \cite{BP04} (before oscillations) is
$131_{-10}^{+12}$. 

This simple fact has highly non-trivial implications. Indeed, this
behavior is not ``generic'' for mass-induced oscillations, even when
they combine
with the MSW \cite{Wolfenstein:1977ue,Mikheev:1986gs} matter effect.
{\it A priori}, one might have expected solar neutrinos to be in one
of these regimes:
\begin{itemize}
\item matter dominates at the production point, on the way out of the
  Sun neutrino flavor evolves adiabatically $\rightarrow$ constant
  suppression (regime 1);
\item matter dominates at the production point, on the way out of the
  Sun neutrino flavor evolves non-adiabatically $\rightarrow$ vacuum
  oscillations
  \begin{itemize}
  \item vacuum oscillation length $\ll$ 1 a.u. (astronomical unit)
    $\rightarrow$ oscillations average out $\rightarrow$ constant
    observed suppression (regime 2);
  \item vacuum oscillation length $\gg$ 1 a.u. (astronomical unit)
    $\rightarrow$ no time to oscillate $\rightarrow$ no suppression
    (regime 3);
  \end{itemize}
\item vacuum oscillations dominate everywhere, matter effects
  negligible even in the center of the Sun $\rightarrow$ oscillations
  average out $\rightarrow$ constant observed suppression (regime 4).
\end{itemize}
The observed energy-dependent $P_{ee}^{std}$ then implies
that solar neutrinos are in one of the several ``special regimes'':
the transition between regimes 1 and 4 (Large Mixing Angle -- LMA --
solution); the transition between regimes 2 and 3 (vacuum/quasi-vacuum
\cite{Friedland:2000cp} oscillation solution); the transition between
regimes 1 and 2 (Small Mixing Angle -- SMA -- solution); the regime
where the density in the Earth is close to resonant, so that the
flavor regeneration in the Earth is large (the LOW solution). These
solutions were known for many years, in particular all four were
allowed as recently as 2000, see, e.g., \cite{deGouvea:2000cq}.

We now of course know that only the LMA solution survives. Let us
consider the survival probability $P_{ee}^{std}$ under the (\textit{a
  posteriori} justified) assumption that the oscillations take place
between just two eigenstates. One easily obtains that during the day time
\begin{equation}\label{eq:Peestd}
    P_{ee}^{std,\;2\nu}=\cos^2\theta_\odot\cos^2\theta+\sin^2\theta_\odot\sin^2\theta.
\end{equation}
The probability of finding the neutrino in eigenstate 1(2) is
$\cos^2\theta_\odot$($\sin^2\theta_\odot$), where $\theta_\odot$
is the mixing angle at the production point; in turn, the
probability of detecting the neutrino already in eigenstate 1(2)
as $\nu_e$ is $\cos^2\theta$($\sin^2\theta$). The key physical
ideas here are that the evolution is adiabatic (no level jumping)
and incoherent (interferences between 1 and 2 disappear upon
integration over energies for $\Delta m^2 \gtrsim 10^{-9}-10^{-8}$
eV$^2$ \cite{Pakvasa:1990gf,Friedland:2000cp} and over the
production region).

The angle $\theta_\odot$ is determined from the oscillation
Hamiltonian $H_{\rm tot}= H_{\rm vac} + H_{\rm mat}$, where
\begin{eqnarray}
H_{\rm vac} = \left(\begin{array}{rr}
  -\Delta \cos 2\theta & \Delta \sin 2\theta \\
  \Delta \sin 2\theta & \Delta \cos 2\theta
\end{array} \right),\;\;\;H_{\rm mat} = \left(\begin{array}{cc}
  \sqrt{2} G_F n_e & 0 \\
  0 & 0
\end{array} \right).
\label{eq:VAC_convention}
\end{eqnarray}
Here $\Delta\equiv \Delta m^2/(4 E_\nu)$ and $\Delta m^2$ is the
mass splitting between the first and second neutrino mass states:
$\Delta m^2\equiv m^2_2-m^2_1$. The two limiting values are
$\theta_\odot=\theta$ ($H_{\rm tot}$ is dominated by $H_{\rm
vac}$) and $\theta_\odot=\pi/2$ ($H_{\rm tot}$ is dominated by
$H_{\rm mat}$). The probability $P_{ee}^{std}$ then varies from
$\cos^4\theta+\sin^4\theta$ ($=1-(1/2)\sin^22\theta$, averaged
vacuum oscillations) to $\sin^2\theta$.

\begin{figure}[htbp]
  \centering
  \includegraphics[width=0.77\textwidth]{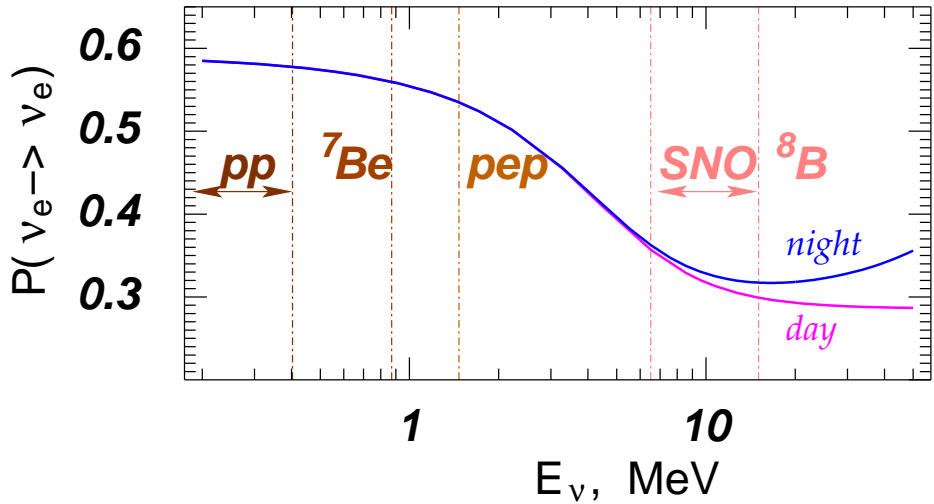}
  \caption{The $\nu_e$ survival probability and day/night
  asymmetry for  the LMA solution.}
  \label{fig:PeeLMA}
\end{figure}

The transition from one regime to another occurs when $\sqrt{2}
G_F n_e \sim 2 \Delta$ at the production point. To accommodate the
data on $P_{ee}$, this transition must occur \emph{right in the
middle of the solar neutrino spectrum}, implying $\Delta m^2\sim
\mbox{a few}\times 10^{-5}$ eV$^2$. Moreover, $\Delta m^2$ cannot
be lower than $\sim3\times10^{-5}$ eV$^2$ to avoid being close to
the resonance condition in the Earth and resulting large day/night
variations of $P_{ee}$. The situation is illustrated in
Fig.~\ref{fig:PeeLMA}.  Evidently, Nature chose to ``tune'' the
mass splitting involved in solar neutrino oscillations to the
density in the solar core! Remarkably, a completely independent
reactor antineutrino measurement by KamLAND
\cite{KamLAND2002flux} showed that $\Delta m^2$ is indeed in this
range.


The preceding discussion assumed that mass eigenstate 3 is not
involved in the evolution of solar neutrinos.
%
The correction due its presence is trivially computed if we notice
that the splitting between this state and eigenstates 1 and 2 is
significantly larger than the matter potential even in the center
of the Sun ($\Delta m_{atm}^2/2E \gg \sqrt{2}G_F N_e(0)$), so that
the $\nu_e$ content of that state is always given by
$\sin^2\theta_{13}$. Repeating the arguments that led to
Eq.~(\ref{eq:Peestd}), one gets
\begin{equation}\label{eq:Peestd3nu}
    P_{ee}^{std,\;3\nu}=\sin^4\theta_{13}+\cos^4\theta_{13}P_{ee}^{std,\;2\nu}.
\end{equation}
Given the bound $\sin^2\theta_{13}\lesssim0.02$ from CHOOZ
\cite{Apollonio:2002gd}, the first term is negligibly small. The
effect of the third state then is to multiply the two-neutrino
survival probability by $\cos^4\theta_{13}$. The resulting
correction is at most 4\%; this correction is basically the
probability that the original electron neutrino ``disappears''
into state 3. See, {\it e.g.}, \cite{Goswami:2004cn,Fogli:2005cq}
for recent data analyses and further references.

\section{Searching for nonstandard neutrino interactions}

The impact of nonstandard neutrino interactions with matter on
solar neutrino oscillations was discussed already in the classical
paper by L. Wolfenstein \cite{Wolfenstein:1977ue} and subsequently
elaborated on by many authors
(\cite{Valle:1987gv,Roulet:1991sm,Guzzo:1991hi} and many others).
The  idea is that novel interactions due to a heavy vector and
scalar exchange could modify the neutrino forward scattering
amplitude and hence the oscillation Hamiltonian in matter.
Regardless of their origin, at low energies relevant to neutrino
oscillations, nonstandard interactions (NSI) are described by the
effective Lagrangian
\begin{eqnarray}
L^{NSI} = - 2\sqrt{2}G_F (\bar{\nu}_\alpha\gamma_\rho\nu_\beta)
(\epsilon_{\alpha\beta}^{f L}\bar{f}_L \gamma^\rho f_L +
\epsilon_{\alpha\beta}^{f R}\bar{f}_R\gamma^\rho f_{R})+ h.c.
\label{eq:lagNSI}
\end{eqnarray}
Here $\epsilon_{\alpha\beta}^{f L}$ ($\epsilon_{\alpha\beta}^{f
R}$) denotes the strength of the NSI between the neutrinos $\nu$
of flavors $\alpha$ and $\beta$ and the left-handed (right-handed)
components of the fermions $f$.

Neutrino scattering tests, like those of NuTeV
\cite{Zeller:2001hh} and CHARM \cite{Vilain:1994qy}, 
constrain mainly the NSI couplings of the muon neutrino, e.g.,
$|\epsilon_{e\mu}|\lesssim 10^{-3}$, $|\epsilon_{\mu\mu}|\lesssim
10^{-3}-10^{-2}$. In contrast, direct limits on $\epsilon_{ee}$,
$\epsilon_{e\tau}$, and $\epsilon_{\tau\tau}$ are remarkably
loose, e. g., $|\epsilon_{\tau\tau}^{uu R}|<3$,
$-0.4<\epsilon_{ee}^{uu
  R}<0.7$, $|\epsilon_{\tau e}^{uu}|<0.5$, $|\epsilon_{\tau
  e}^{dd}|<0.5 $ \cite{Davidson:2003ha}. Stronger constraints exist on
the corresponding interactions involving the charged leptons.
Those, however, are model-dependent and do not apply if the NSI
come from the underlying operators containing the Higgs fields
\cite{Berezhiani:2001rs}. Here we only consider direct
experimental bounds.

Even with the addition of NSI the splitting $\Delta m_{atm}^2/2E$
remains much greater than the matter potential anywhere along the
neutrino trajectory. This means the solar neutrino analysis can
still be reduced to two neutrino states, following the arguments
of Sect.~\ref{sect:stdLMA}.
Neglecting small corrections of order $\sin \theta_{13}$ or
higher, the corresponding matter contribution to the two-neutrino
oscillation Hamiltonian can be written as
\begin{eqnarray}
H_{\rm mat}^{NSI} = \frac{G_F n_e}{\sqrt{2}}
\left(\begin{array}{cc}
   1+\epsilon_{11} & \epsilon^\ast_{12} \\
  \epsilon_{12}   & -1-\epsilon_{11}
\end{array} \right), \mbox{  where  }
\begin{array}{l}
\epsilon_{11}=\epsilon_{ee} - \epsilon_{\tau\tau}
\sin^2\theta_{23}, \\
 \epsilon_{12}=-2\epsilon_{e\tau} \sin
\theta_{23}.
\end{array}
\label{eq:ceciconv}
\end{eqnarray}
The epsilons are the sums of the contributions from the matter
constituents: $\epsilon_{\alpha\beta}\equiv
\sum_{f=u,d,e}\epsilon_{\alpha\beta}^{f}n_f/n_e$. In turn,
$\epsilon_{\alpha\beta}^{f}\equiv\epsilon_{\alpha\beta}^{fL}+\epsilon_{\alpha\beta}^{fR}$.
Observe that only the vector component of the NSI enters the
propagation effect; in contrast, the NC detection process at SNO
depends on the axial coupling. The propagation and detection
effects of the NSI are thus sensitive to different parameters, and
the corresponding searches could be complementary.

Eq.~(\ref{eq:ceciconv}) shows that the flavor changing NSI effect
in solar neutrino oscillations comes from $\epsilon_{e\tau}$,
while the flavor preserving NSI effect comes from both
$\epsilon_{ee}$ and $\epsilon_{\tau\tau}$. A useful
parameterization is
\begin{eqnarray}
\label{eq:MAT_convention}
 H_{\rm mat}^{NSI} =
\left(\begin{array}{cc}
  A \cos 2\alpha & A e^{-2i\phi} \sin 2\alpha \\
  A e^{2i\phi} \sin 2\alpha  & -A \cos 2\alpha
\end{array} \right),
 \mbox{  where  }
\begin{array}{l}
\tan 2\alpha = |\epsilon_{12}|/(1+\epsilon_{11}) ,\\
2\phi=Arg(\epsilon_{12}), \\
A= G_F n_e \sqrt{[(1+\epsilon_{11})^2+|\epsilon_{12}|^2]/2}~.
\end{array}
\end{eqnarray}
The effect of $\alpha$ is to change the mixing angle in the
medium of high density from $\pi/2$ to $\pi/2-\alpha$. The angle
$\phi$ (related to the phase of $\epsilon_{e\tau}$) is a source of
CP violation. Solar neutrino experiments, just like terrestrial
beam experiments \cite{Gonzalez-Garcia:2001mp,Campanelli:2002cc},
are sensitive to its effects \cite{Friedland:2004pp}, while
atmospheric neutrinos are not
  \cite{Friedland:2004ah,Friedland:2005vy}.

\begin{figure}[htbp]
  \centering
  \includegraphics[width=0.97\textwidth]{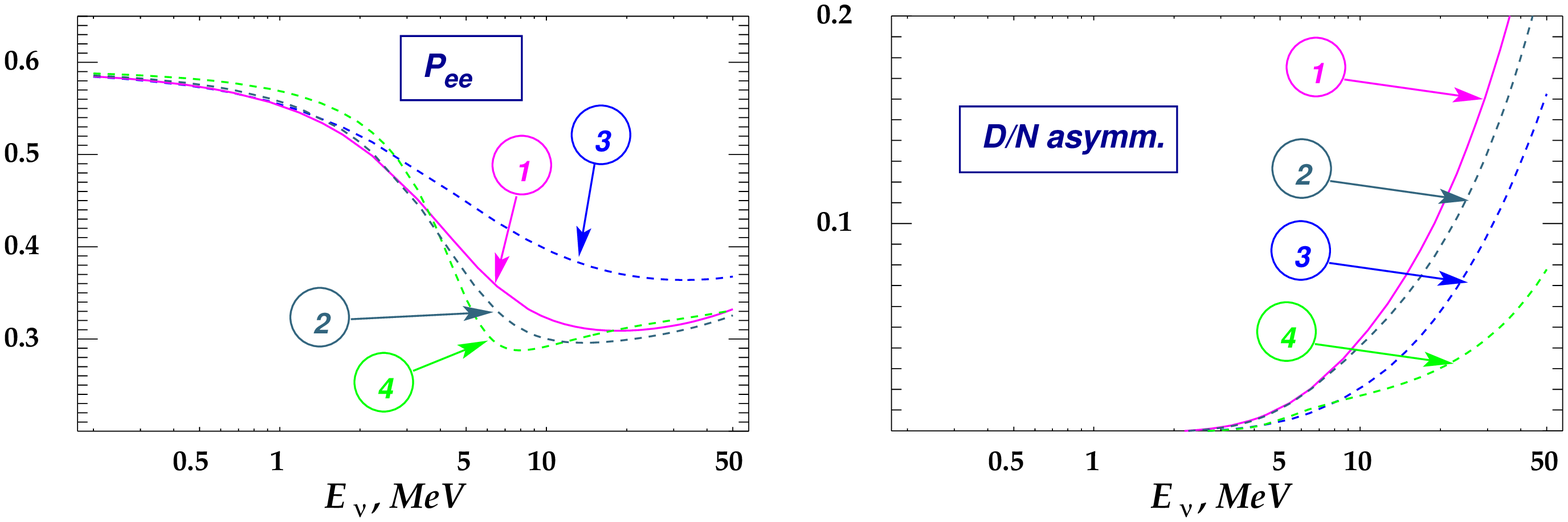}
  \caption{The electron neutrino survival probability (\emph{left}) and day/night
  asymmetry (\emph{right})
  for  $\Delta m^2=7\times 10^{-5}$ eV$^2$,
  $\tan^2\theta=0.4$ and several representative values of the NSI parameters: (1)
  $\epsilon_{11}^{u}=\epsilon_{11}^{d}=\epsilon_{12}^{u}=\epsilon_{12}^{d}=0$;
  (2) $\epsilon_{11}^{u}=\epsilon_{11}^{d}=-0.008$,
  $\epsilon_{12}^{u}=\epsilon_{12}^{d}=-0.06$;
  (3) $\epsilon_{11}^{u}=\epsilon_{11}^{d}=-0.044$,
  $\epsilon_{12}^{u}=\epsilon_{12}^{d}=0.14$;
  (4) $\epsilon_{11}^{u}=\epsilon_{11}^{d}=-0.044$,
  $\epsilon_{12}^{u}=\epsilon_{12}^{d}=-0.14$.
  Reproduced from \protect\cite{Friedland:2004pp}.}
  \label{fig:PeeAdn}
\end{figure}

The main effects of NSI on $P_{ee}$ are as follows
\cite{Friedland:2004pp}: (i) the low-energy limit stays the same
(vacuum oscillations); (ii) the high-energy limit changes,
according to Eq.~(\ref{eq:Peestd}), $P_{ee}\rightarrow
\sin^2\alpha\cos^2\theta+\cos^2\alpha\sin^2\theta$; (iii) at
intermediate energies, the transition from vacuum to matter
dominated regime can shift in energy, with changing $A$, and can
become more or less abrupt, with changing $\alpha$ and $\phi$. The
nonadiabatic regime occurs when $\theta\rightarrow\alpha$, rather
than $\theta\rightarrow0$. Also, very importantly, the day/night
effect can change with \emph{all three parameters}. In particular,
it becomes small either as $A\rightarrow0$
\cite{Guzzo:2004ue,Miranda:2004nb} or as $\alpha\rightarrow\theta$
\cite{Friedland:2004pp}. Thus, the LMA-0 region that is normally
excluded by the non-observation of day/night asymmetry may become
allowed \cite{Friedland:2004pp,Guzzo:2004ue,Miranda:2004nb}.

Fig.~\ref{fig:PeeAdn} illustrates the impact of the NSI on
$P_{ee}$ and the day/night asymmetry. Curve 3 gives an example of
parameters that can already be excluded by the current data. Curve
4 illustrates the suppression of the Earth effect described above.
For technical details, including approximate analytical
expressions for $P_{ee}$ and day/night asymmetry, see
\cite{Friedland:2004pp}.

\begin{figure}[htbp]
  \centering
  \includegraphics[width=0.45\textwidth]{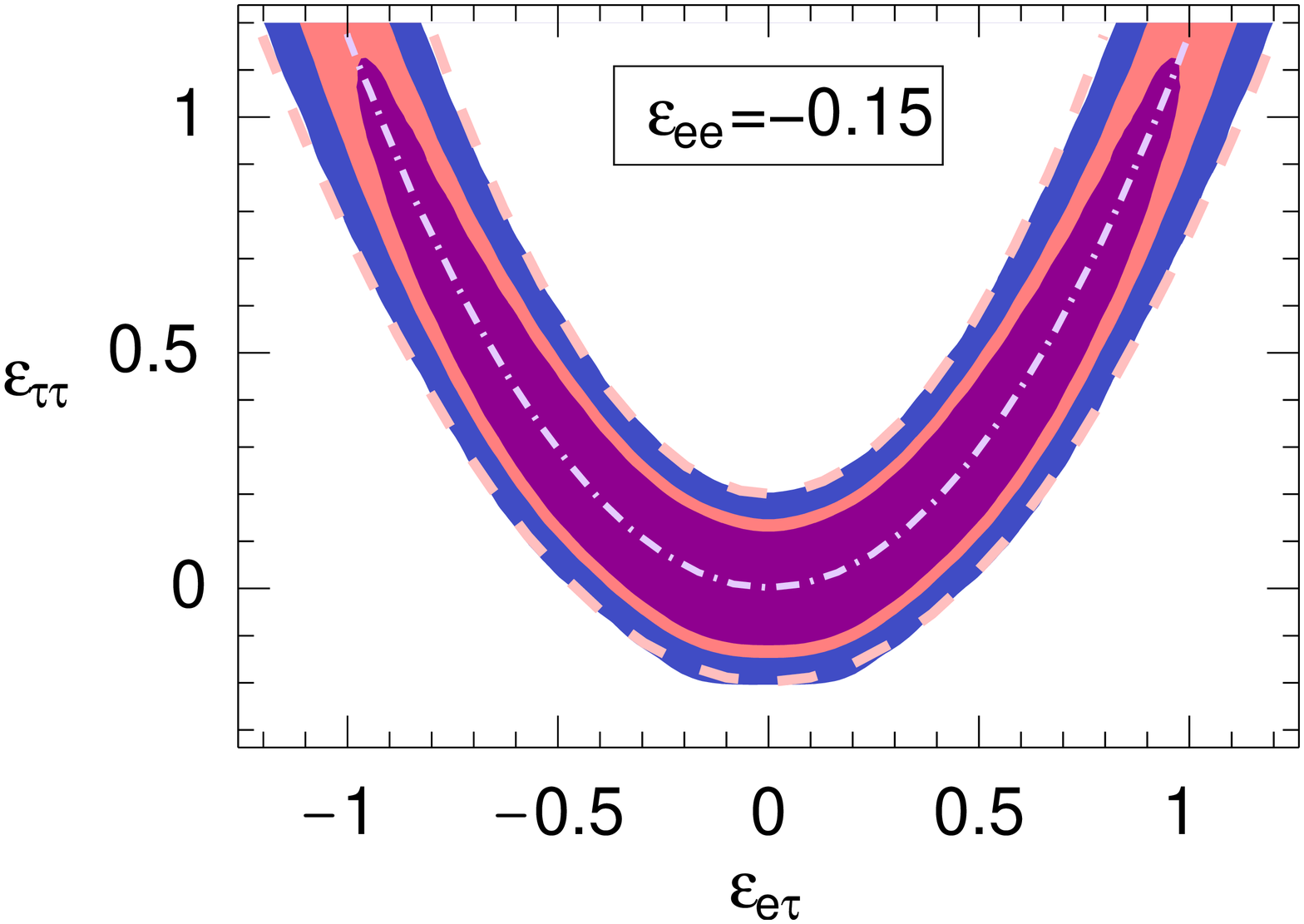}
  \includegraphics[width=0.49\textwidth]{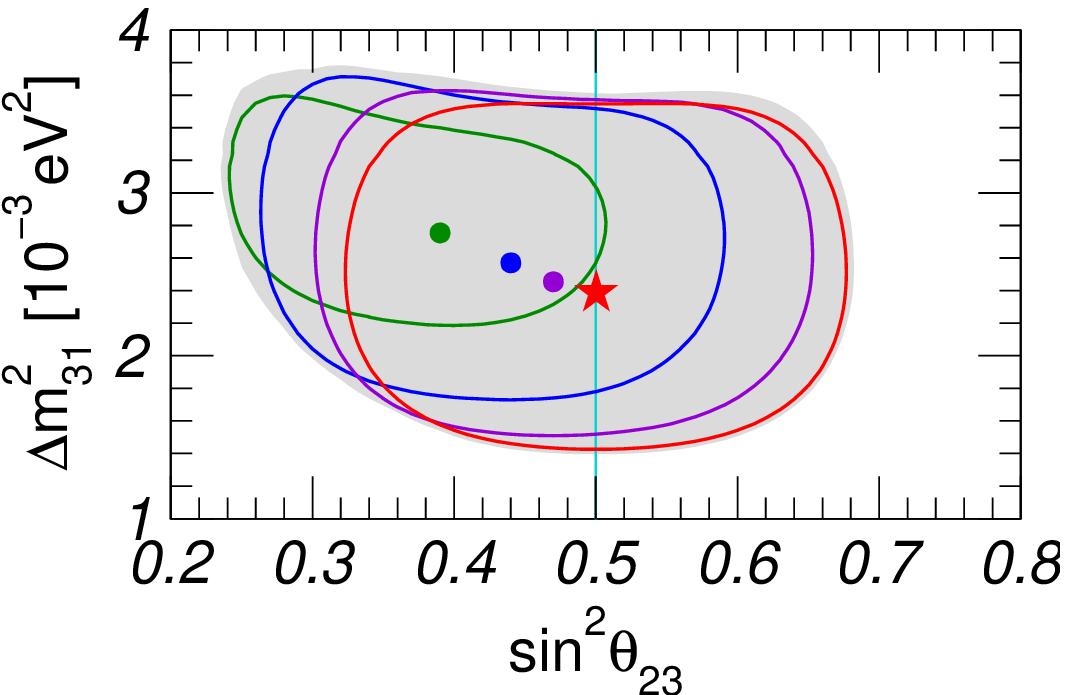}
  \caption{\textit{Left panel}: A 2-D section $(\epsilon_{ee}=-0.15)$ of the allowed region
  of the NSI parameters (shaded).
  We assumed $\Delta m^2_\odot=0$ and
  $\theta_{13}=0$, and marginalized over $\theta$ and $\Delta m^2 $.
  The dashed contours indicate our analytical predictions. See text for
  details. \textit{Right panel}: The effect of the NSI on the allowed region and best-fit values of the
  oscillation parameters. From
  \protect\cite{Friedland:2004ah}.
  }
  \label{fig:scan}
\end{figure}

The solar neutrino analysis of NSI cannot be done in isolation:
the same NSI can also be probed with atmospheric neutrinos.
Indeed, on general grounds, one expects the atmospheric neutrinos
-- particularly the high energy ones for which nonstandard matter
effects can dominate over the vacuum oscillation effects -- to be
a very sensitive probe of NSI. Early two-neutrino
($\nu_\mu,\nu_\tau$) numerical studies \cite{Maltoni2001} yielded
$\epsilon_{\mu\tau}\lesssim 0.08-0.12$ and
$\epsilon_{\tau\tau}\lesssim 0.2$ \footnote{Notice the difference
  in normalization: our $\epsilon$'s are normalized per electron, while
  \cite{Maltoni2001} gives $\epsilon$'s per $d$ quark.}.
Clearly, these are very strong bounds; if they were to extend to
$\epsilon_{e\tau}$, the large NSI effects on solar neutrinos discussed
above would be excluded. It turns out, however, that this is not the
case: when the analysis is properly extended to three flavors, one
finds that very large values of both $\epsilon_{e\tau}$ and
$\epsilon_{\tau\tau}$ are still allowed by the data
\cite{Friedland:2004ah}. This is illustrated in Fig.~\ref{fig:scan}
(\emph{left panel}), which shows that NSI with strengths comparable to
the Standard Model interactions can be compatible with all atmospheric
data. It must be noted that the compatibility is achieved as a result
of adjusting the vacuum oscillation parameters: large NSI imply a
smaller mixing angle and larger $\Delta m_{atm}^2$, as can be see in
the \emph{right panel} of Fig.~\ref{fig:scan}.

The addition of the K2K data helps constrain the allowed NSI
region somewhat \cite{Friedland:2005vy}. While the addition of the
first data  from MINOS brings no further improvement
\cite{Friedland:2006pi}, the future high-statistics MINOS dataset
will be a very valuable probe of this parameter space
\cite{Friedland:2006pi}.

\section{Searching for neutrino transition moments}

The idea that solar neutrinos could be affected by the neutrino
spin precession (NSP) in the solar magnetic fields is even older
\cite{Cisneros:1970nq} than the NSI idea. Remarkably, this idea --
much improved with time
\cite{Voloshin:1986ty,Okun1986short,Okun1986,Akhmedov1988,LimMarciano,Raghavan}
-- remained viable for the next three decades. While, by the late
1990's, the lack of time variations in the Super-Kamiokande data
gave strong evidence against large NSP in the solar convective zone, NSP
in the radiative zone continued to give a good fit to all solar
data \cite{ourmagnfit}.

\begin{figure*}[htbp]
  \centering
  \includegraphics[width=0.77\textwidth]{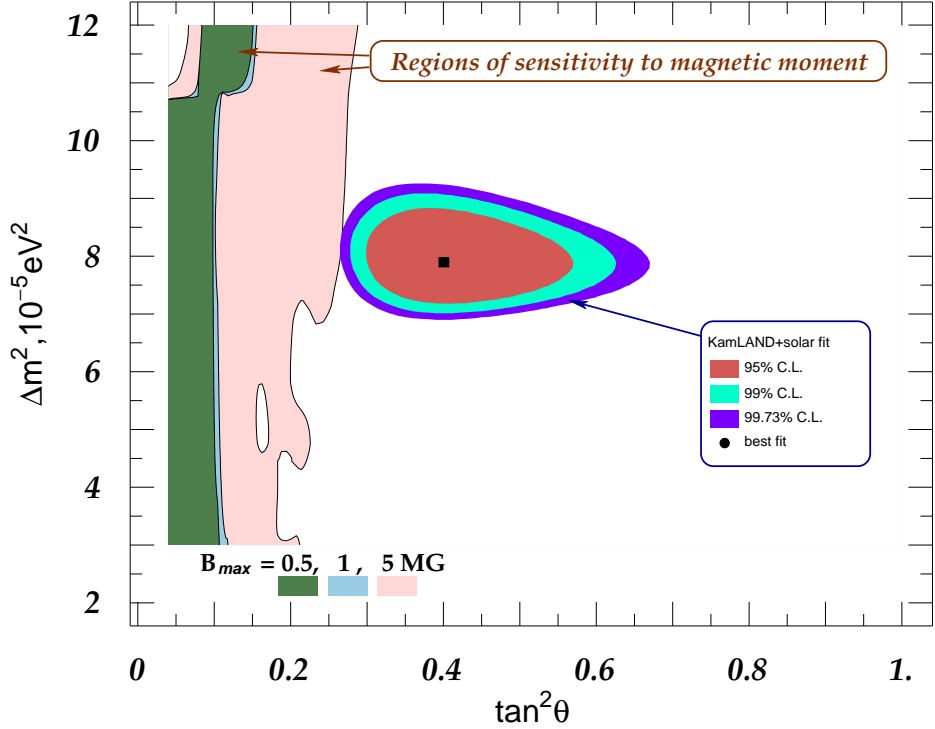}
  \caption{The regions of the oscillation parameter space where
    one may expect the electron antineutrino flux above the KamLAND
    bound \cite{Eguchi:2003gg} (three different shadings
    correspond to three different normalizations of the magnetic
    field, up to the upper bound \cite{magneticbound}). An optimistic value of the transition moment, $\mu=
    1\times 10^{-11}\mu_B$, was taken. For comparison, the region
    allowed by the combined analysis of the KamLAND and solar neutrino
    data \cite{KamLANDspectrum} is also shown. From \protect\cite{Friedland:2005xh}.}
  \label{fig:scan_magn}
\end{figure*}

Even after the confirmation of the LMA oscillation solution by
KamLAND \cite{KamLAND2002flux}, the possibility of the NSP
happening at a \emph{subdominant} level remains of great interest,
as a probe of the neutrino electromagnetic properties and, at the
same time, of the magnetic fields in the solar interior. NSP
coupled with flavor oscillations could lead to conversion
$\nu_e\rightarrow\bar\nu_e$ in the Sun, on which recently, KamLAND
\cite{Eguchi:2003gg} reported an upper bound. It is very important
to understand what this bound implies for the neutrino magnetic
(transition) moment and how it compares with other available bounds on
the neutrino transition moment.

The laboratory bounds on the neutrino magnetic (transition) moment
come from measuring the cross sections of $\nu e^-$ or $\bar\nu e^-$
scattering in nearly forward direction. The recent bound for the
interaction involving the Majorana electron antineutrino is
$2\mu_{e\beta} < 0.9 \times 10^{-10}\mu_B$ at the 90\% confidence
level \cite{munu2005}, where $\mu_B \equiv e/(2 m_e)$ is the Bohr
magneton ($m_e$ is the electron mass, $e$ is its charge). Stronger
bounds, $\mu \lesssim 3 \times 10^{-12}\mu_B$, exist from
astrophysical considerations, particularly from the study of red giant
populations in globular clusters \cite{Raffeltbound}. Larger values of
the transition moment would provide an additional cooling mechanism --
via plasmon decay to $\nu\bar\nu$ -- for the red giant core and change
the core mass at helium flash beyond what is observationally allowed.

An important theoretical consideration is that the magnetic moment
operators will radiatively generate the neutrino mass. Requiring that
the corresponding contribution to the neutrino masses be not much
greater than their observed values, one obtains a model-independent
``naturalness" bound on $\mu_{ab}$. For Dirac neutrinos, the bound
obtained in this way turns out to be very stringent,
$\mu_{ab}\lesssim10^{-14}\mu_B$ \cite{Bell:2005kz}.  However, very
importantly, for Majorana neutrinos the bound is much weaker, only
$\mu_{ab}\lesssim10^{-10}\mu_B$ \cite{Davidson:2005cs,Bell:2006wi},
owing to the different flavor symmetry properties of the mass and the
transition moment operators \cite{Voloshin:1987qy}. In fact, explicit
models exploiting these very symmetry properties were discussed many
years ago ({\it e.g.}, \cite{BabuMohapatra,BabuMohapatra1990}
\footnote{I thank R.~N.~Mohapatra for bringing the last two references
  to my attention.}).

It turns out that \emph{for the measured LMA oscillation
parameters} NSP in the radiative zone cannot produce the
$\bar\nu_e$ flux above the KamLAND bound. This is illustrated in
Fig.~\ref{fig:scan_magn}. This is a remarkable example that
knowing neutrino oscillation parameters precisely can be very
valuable: the answer would qualitatively change if the mixing
angle were 20$^\circ$ instead of 30$^\circ$.

For NSP in the convective zone, the analysis is very different, though
in the end the conclusion is similar: one should not have expected the
flux of $\bar\nu_e$ in excess of the published KamLAND bound. Put
another way, the bound on the neutrino transition moment from the
KamLAND bound is comparable to the direct laboratory and
``naturalness'' bounds, but still weaker than that from analysis of
the red giant cooling. An updated analysis of a larger KamLAND dataset
is needed. The reader is referred to \cite{Friedland:2005xh} for
details and further references.

\ack

It is a great pleasure to acknowledge my collaborators -- C.
Lunardini, M. Maltoni and C. Pe\~na-Garay. I owe special thanks to A.
Gruzinov for countless -- always very clear and helpful -- discussions
of the solar magnetic fields and plasma physics in general. I also
thank M. Rempel for a very helpful discussion and for pointing me to
an excellent set of references. I benefited greatly from stimulating
conversations with V.~Cirigliano, E.  Akhmedov and T. Rashba. Finally,
I thank R. Mohapatra for bringing to my attention several important
references.

\bibliographystyle{jhep}
\bibliography{friedland_neutrino06}

\end{document}